\documentclass[preprint2,flushrt]{aastex}

\usepackage{amsmath}
\usepackage{natbib}
\bibpunct{(}{)}{;}{a}{}{,}
\begin{document}

\title{Acceleration of Energetic Particles Through Self-Generated Waves in a Decelerating Coronal Shock}

\keywords{coronal shock particle acceleration, decelerating shock \\\emph{PACS:} 96.50.Pw 96.50.Tf 96.60.ph}

\author{M. Battarbee\altaffilmark{1}, T. Laitinen\altaffilmark{1}, R. Vainio\altaffilmark{2} and N. Agueda\altaffilmark{2,3}}
\altaffiltext{1}{Department of Physics and Astronomy, University of Turku, Finland; \textbf{markus.battarbee@utu.fi}}
\altaffiltext{2}{Department of Physics, University of Helsinki, Finland}
\altaffiltext{3}{Now at the Space Sciences Laboratory, University of California, Berkeley, USA} 

\begin{abstract}
We have developed a simulation model of particle acceleration in
coronal shock waves. The model is based on a Monte Carlo method, where
particles are traced in prescribed large-scale electromagnetic fields
utilizing the guiding center approximation.  The particles are
scattered in the turbulence according to quasilinear theory, with the
scattering amplitude directly proportional to the intensity of
Alfv\'en waves at gyro-resonant wavenumbers.  The Alfv\'en waves are
traced simultaneously with the particles, so that the wave field is
propagated outwards from the Sun using WKB propagation supplemented
with a phenomenological wavenumber diffusion term and a growth rate
computed from the net flux of the accelerated particles.  We consider
initial wave amplitudes small enough to allow rapid escape of
particles from the shock to the ambient medium. Thus, in our model the
Alfv\'en waves responsible for the diffusive acceleration of particles
are generated by the accelerated particles themselves.  In this work,
we study the effects of non-constant shock velocity and non-monotonic
Alfv\'en velocity on particle acceleration scenarios.  We report in
particular how the deceleration of a shock affects particle intensity
and turbulence power evolution in the vicinity of the shock.
\end{abstract}

\section{Introduction}

Acceleration by coronal and interplanetary shocks driven by coronal mass ejections (CMEs) is currently considered the primary source of large solar energetic particle (SEP) intensities. Injected particles repeatedly cross the shock front, scattering from the plasma waves, and gain energy on each crossing. A requirement for rapid acceleration and escape is strong turbulent trapping in front of the shock and weaker trapping further out. Such a turbulent structure can, in large events, be maintained by streaming of the accelerated particles themselves, as plasma waves in the upstream experience net amplification by scattering events. Diffusive acceleration of solar energetic particles in coronal shocks driven by fast CMEs has been discussed by eg. \cite{1977ICRC...11..132A}; \cite{1978ApJ...221L..29B} and \cite{1977DoSSR.234R1306K}, whereas upstream wave amplification was detailed by \cite{1978MNRAS.182..147B}.

The focus of recent research has been time-dependent numerical modeling. \cite{2007ApJ...658..622V} reported on an approach using a simplified solar wind model utilizing constant wave group speed $V = u_{sw} + v_A$ and constant shock velocity.  The initial study, while bringing new insights into the energetic particle research, could not directly be extended to variable Alfv\'enic speeds as the model used a Lagrangian grid with temporally constant spacing to simplify wave transport modelling.

In this study, we use a shock-attached wave grid, which allows realistic coronal velocity profiles and greater numerical accuracy near the shock. In addition, eliminating interpolation calculations at the shock front stop numerical oscillations from forming. Our model will allow us to implement multiple (in- and outbound) wave modes in future studies. Spatial convection of wave power within the grid was done with a Van Leer flux-limited Lax-Wendroff scheme \citep{vanLeer1974361,LW60}.

\section{model}
\subsection{Wave-particle interactions}

In our coupled SEP acceleration and Alfv\'en wave self-generation model \citep{2007ApJ...658..622V}, the particles are simulated with a Monte-Carlo method, while the waves are described with a frequency power spectrum, propagated in the plasma background with the WKB wave transport equation supplemented by spectral transport and amplification terms,
\begin{equation}
\frac{\partial\tilde{P}}{\partial t} +v\frac{\partial\tilde{P}}{\partial r} = \sigma \tilde{P} +\frac{\partial}{\partial f}\left(D_{ff}\frac{\partial\tilde{P}}{\partial f}\right).\nonumber
\end{equation}
We use a simplified form of the amplification coefficient
\begin{equation}
\sigma(f,r,t)=\pi^{2}f_{cp} \frac{p S_{p}(r,t)}{nv_{A}}\nonumber
\end{equation}
as derived in \cite{2003A&A...406..735V}, where 
\begin{equation}
S_{p}(r,t)=2\pi p^{2}v \int^{+1}_{-1} \mu F(r,p,\mu,t)d\mu\nonumber
\end{equation}
represents the particle streaming in the Alfv\'en wave frame. For the wave diffusion coefficient, we use a linear diffusion model as given in \cite{2007ApJ...658..622V},
\begin{equation}
D_{ff} = (V/r_{\oplus})f^{8/3}f_{b}^{-2/3}.\nonumber
\end{equation}
In the above equations, $\tilde{P}$ is the normalized wave power $\tilde{P}\equiv(V^{2}/Bv_{A})P$, $f_{cp}$ is the proton cyclotron frequency and $p=m_{p}Vf_{cp}/f$ is the resonant wave-frame momentum simplified by neglecting its dependence on particle pitch angle. The wave group speed is $V$, $r_{\oplus}$ = 1 AU, $F$ is the proton distribution function, and the breakpoint frequency $f_{b}$ is set to 1 mHz to match detected spectra at 1 AU. Streaming is tracked via particle movement between simulation cells, weighted to take into account the movement of the grid. Wave-particle interactions are implemented so that particle energies can be relativistic. Particles are not traced downstream of the shock, but rather instantly returned upstream with their statistical weight adjusted by the theoretical return probability.

\subsection{Solar wind model}

Our solar wind model consists of a background solar wind with speed $u(r)$, electron density $n(r)$ and a super-radially expanding coronal/interplanetary magnetic flux tube. The magnetic field at the axis of the flux tube is
\begin{equation}
B(r)=B_0\left(\frac{r_\oplus}{r}\right)^2\left[1+1.9\left(\frac{R_\odot}{r}\right)^6\right],\nonumber
\end{equation}
with $B_0$ scaling the values to 2.90 nT at 1 AU \citep{2003A&A...407..713V}.
The cross-section of the flux tube is inversely proportional to the magnetic field, and the off-axis component of the magnetic field, while satisfying the $\nabla \cdot \vec B=0$ condition everywhere, vanishes on the axis. The parallel shock is taken to propagate at speed $V_s(r)$ along the mean magnetic field.
For the electron density, we use the empirical model
in \cite{2005ApJS..156..265C},
\begin{align}
n(r_{n})=n_0 ( 1.0~r_{n}^{-2} +25.0 ~r_{n}^{-4} +300.0 ~r_{n}^{-8} \nonumber\\
 +1500.0 ~r_{n}^{-16} +5796.0 ~r_{n}^{-33.9}),\nonumber
\end{align}
with $r_{n}=r/R_{\odot}$ and $n_0$ set to facilitate, with the rest of the solar wind model, an Alfv\'en speed of 20 km/s at 1 AU.
The solar wind velocity is inferred from mass conservation.
This solar wind model results in a local maximum in the Alfv\'en velocity below $5 R_\odot$.
\begin{figure}[t]
\resizebox{1.0\columnwidth}{!}
 {\includegraphics{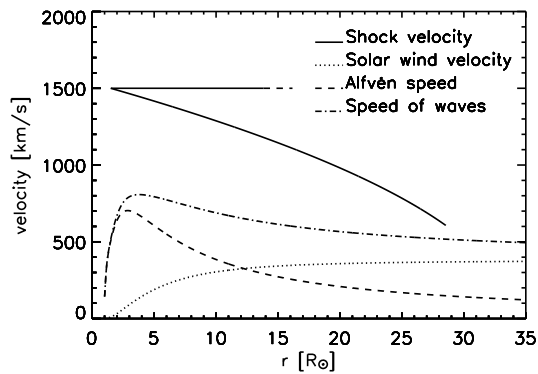}}
  \caption{Solar wind and shock parameters; the line for the constant velocity shock speed is truncated so it does not overlap with the legend.}
\end{figure}

\subsection{Simulation parameters}

In this study, we propagate shocks from 1.5 $R_\odot$ with an initial velocity of 1500 km/s. Turbulence is tracked on a logarithmic grid reaching out to 300 $R_\odot$ in front of the shock. The mean magnetic field is normal to the plane of the shock, and the scattering center compression ratio is set to 4. The ambient 100 keV proton mean free path is set to $\lambda_0=1 R_\odot$ at $r_0 = 1.5 R_\odot$. The total amount of injected particles (within a flux tube of 1 sr solid angle at the solar surface)
is given values of $2\cdot10^{35}$ (run A), $6\cdot10^{35}$ (run B) and $2\cdot10^{36}$ (run C). The injection of a slightly suprathermal particle population is detailed in \cite{2007ApJ...658..622V}. The shocks are propagated with both constant velocity and with a constant deceleration of 50 m/s$^2$. 

\section{Results and discussion}

Our simulations show that both constant velocity and decelerating shocks are capable of significant acceleration of energetic particles, although constant velocity shocks do attain greater maximum energies. For run A, the gradually decreasing shock velocity proves to be significantly less effective in accelerating particles. For run B, the decelerating scenario retains greater turbulence in front of the shock, which results in stronger trapping and continued acceleration. This simulation displays a continuous proton power-law spectrum up to 1 MeV, whereas the constant velocity case retains less high-energy particles near the shock. The increased acceleration efficiency and resulting turbulence generation of case C causes the trapping difference to become insignificant, with both constant velocity and decelerating shocks displaying a continuous proton power-law spectrum up to 20 MeV.

\subsection{Comparison at set propagation distance}

\begin{figure}[tb]
\resizebox{1.0\columnwidth}{!}
 {\includegraphics{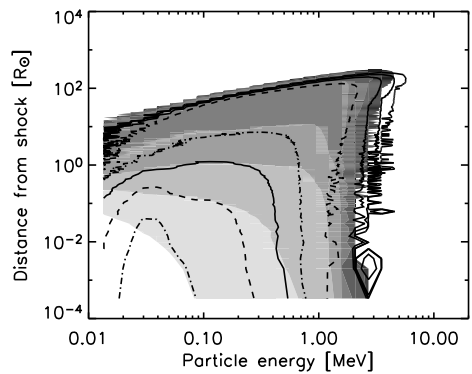}}
  \caption{Proton intensities with injection of $6\cdot10^{35}$ particles (run B), when $r_{s} = 22.7~R_{\odot}$. 
The constant velocity shock values are represented with contour lines, whereas the decelerating case values are presented with filled contours.
Contour lines are at one magnitude intervals.}
\label{6e35_22rs_p}
\end{figure}

\begin{figure}[tb]
\resizebox{1.0\columnwidth}{!}
 {\includegraphics{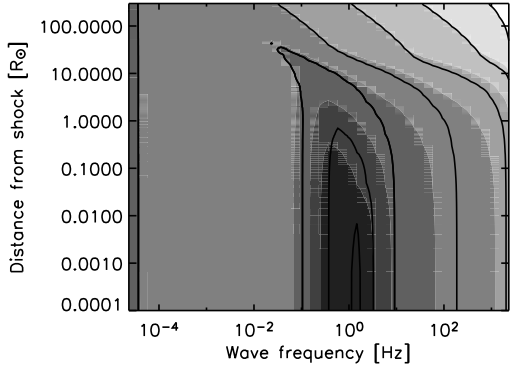}}
  \caption{Wave powers with injection of $6\cdot10^{35}$ particles (run B), when $r_{s} = 22.7~R_{\odot}$. For contour descriptions, see fig.\ref{6e35_22rs_p}}
\label{6e35_22rs_w}
\end{figure}

The particle and wave turbulence distributions (figures \ref{6e35_22rs_p} and \ref{6e35_22rs_w}, respectively) resulting from run B are presented when both shocks have travelled the set distance of 22.7~$R_{\odot}$.

Particles from the decelerating shock have travelled further due to greater total time simulated, but have not reached as high energies as those accelerated by a constant velocity shock. The enhancement in turbulence in front of the decelerating shock is stronger and more diffuse than that of the constant velocity shock, which can be explained by prolonged exposure to efficient wave generation at lower altitudes, and slower convection of the upstream plasma to the shock.


\subsection{Comparison at set propagation time}

\begin{figure}[tb]
\resizebox{1.0\columnwidth}{!}
 {\includegraphics{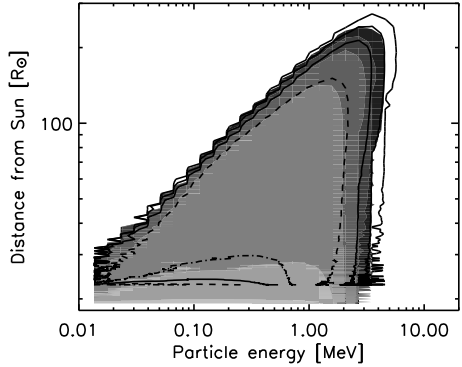}}
  \caption{Proton intensities with injection of $6\cdot10^{35}$ particles (run B), when $t$ = 164 minutes. For contour descriptions, see fig.\ref{6e35_22rs_p}}
\label{6e35_164min_p}
\end{figure}

\begin{figure}[tb]
\resizebox{1.0\columnwidth}{!}
 {\includegraphics{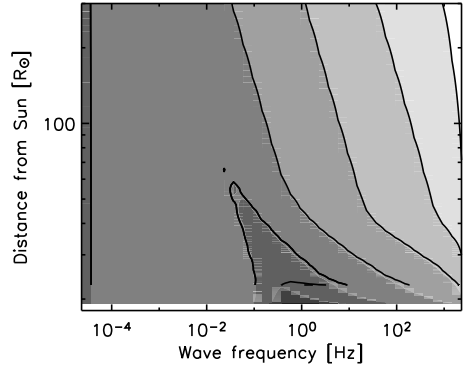}}
  \caption{Wave powers with injection of $6\cdot10^{35}$ particles (run B), when $t$ = 164 minutes. For contour descriptions, see fig.\ref{6e35_22rs_p}}
\label{6e35_164min_w}
\end{figure}

The particle and wave turbulence distributions (figures \ref{6e35_164min_p} and \ref{6e35_164min_w}, respectively) resulting from run B are presented when both shocks have travelled the set time of 164 minutes. Clearly, the constant velocity shock accelerates particles to higher maximum energies and the particles travel further.


\subsection{Spectra in front of the shocks}

\begin{figure}[!tb]
\resizebox{1.0\columnwidth}{!}
 {\includegraphics{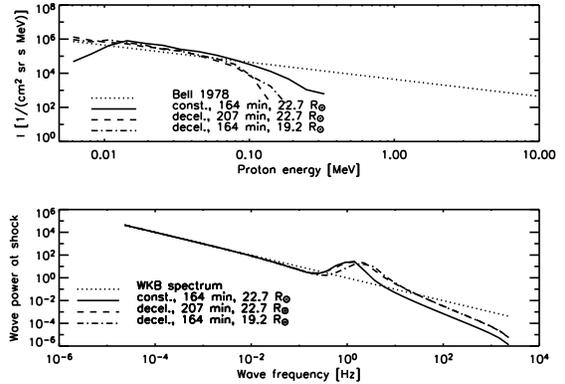}}
  \caption{Proton intensities and wave powers in front of the shock, injection of $2\cdot10^{35}$ particles (run A).}\label{2e35_spect_c}
\end{figure}

\begin{figure}[!tb]
\resizebox{1.0\columnwidth}{!}
 {\includegraphics{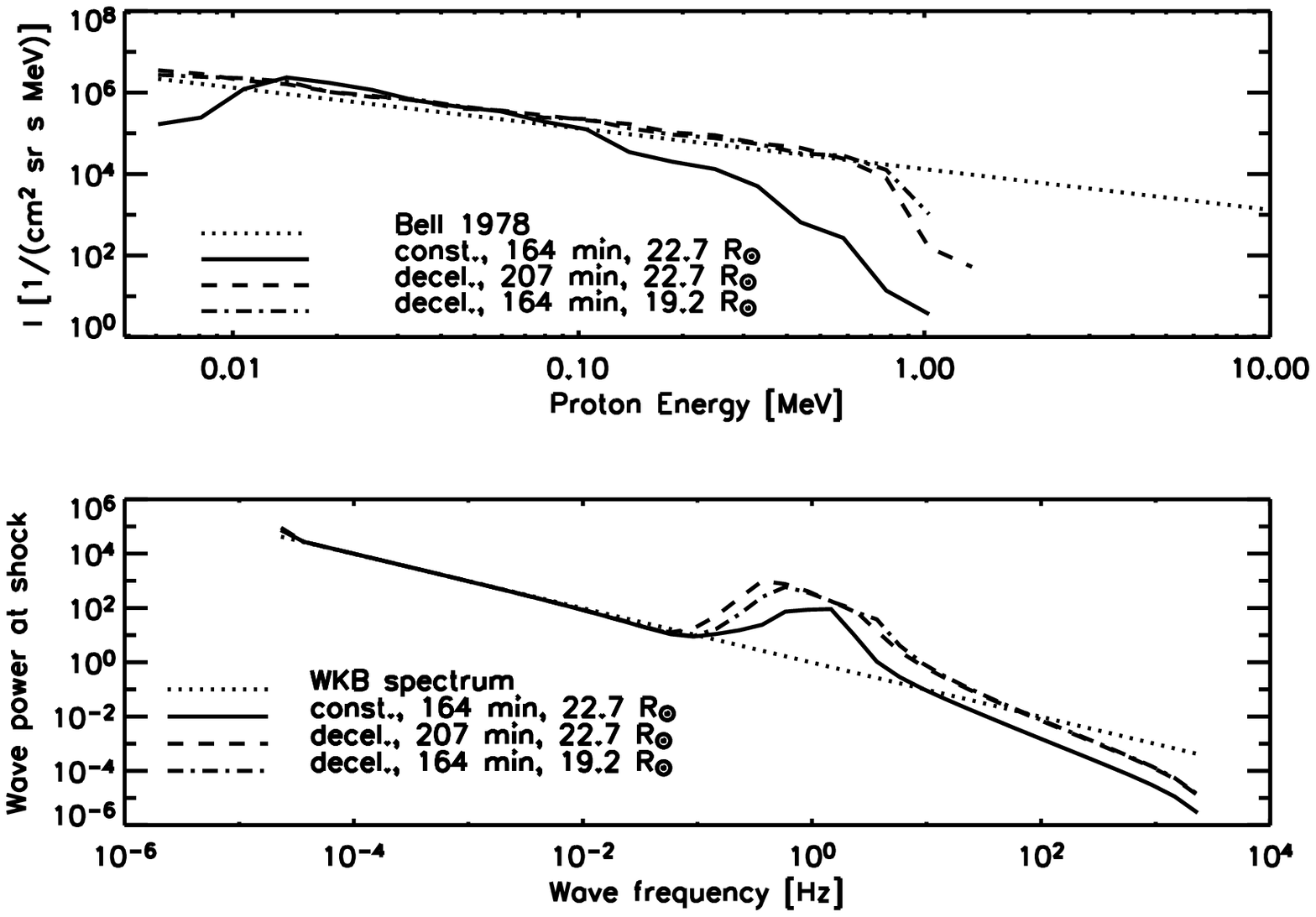}}
  \caption{Proton intensities and wave powers in front of the shock, injection of $6\cdot10^{35}$ particles (run B).}\label{6e35_spect_c}
\end{figure}

\begin{figure}[!tb]
\resizebox{1.0\columnwidth}{!}
 {\includegraphics{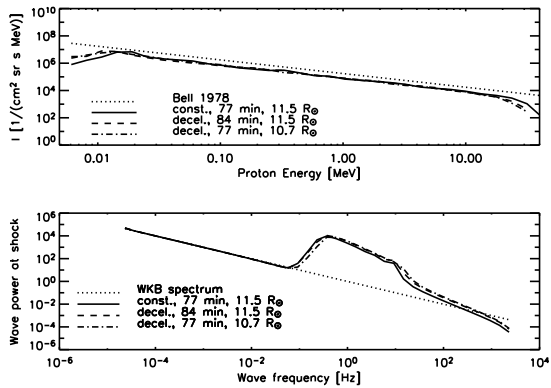}}
  \caption{Proton intensities and wave powers in front of the shock, injection of $2\cdot10^{36}$ particles (run C).}\label{2e36_spect_c}
\end{figure}

Figures \ref{2e35_spect_c}, \ref{6e35_spect_c} and \ref{2e36_spect_c} show the energetic particle intensities and wave power spectra found in front of the shocks for runs A, B and C, respectively. Each figure displays both the spectrum for a constant velocity shock and spectra for decelerating shocks, when the decelerating shock has either travelled the same time or the same distance as the constant velocity shock.
 For comparison, we also present the spectrum according to the steady-state model of \cite{1978MNRAS.182..147B}, as further explained in \cite{2007ApJ...658..622V}. The upstream plasma velocities (927 and 517 km/s) are calculated from the solar wind and shock velocities at the highest altitudes attained  (11.5 $R_{\odot}$ and 22.7 $R_{\odot}$, respectively). It should be noted, that the upstream plasma velocity does not remain constant in our simulations as they do not represent a steady-state scenario.


\section{Conclusions}

Our results suggest that decelerating shocks can accelerate particles to high energies, though not as efficiently as in the case of a constant velocity shock. The acceleration efficiency and the nature of particle trapping vary greatly based on the number of injected particles, which proves that further parameter studies are required to shed light on the dynamical, non-linear effects surrounding decelerating shocks. In addition, the effects of the realistic evolution of the scattering center compression ratio and the particle injection profile based on the solar wind density remain an open question.

\acknowledgments
  The authors would like to thank the IT Center for Science Ltd (CSC) for computational services and the Academy of Finland (AF) for financial support of projects 122041, 121650 and 124837.

\bibliographystyle{aa}

\bibliography{/home/marbat/marbat}

\begin{thebibliography}{10}
\expandafter\ifx\csname natexlab\endcsname\relax\def\natexlab#1{#1}\fi

\bibitem[{{Axford} {et~al.}(1977){Axford}, {Leer}, \&
  {Skadron}}]{1977ICRC...11..132A}
{Axford}, W.~I., {Leer}, E., \& {Skadron}, G. 1977, in International Cosmic Ray
  Conference, Vol.~11, International Cosmic Ray Conference, 132--+

\bibitem[{{Bell}(1978)}]{1978MNRAS.182..147B}
{Bell}, A.~R. 1978, \mnras, 182, 147

\bibitem[{{Blandford} \& {Ostriker}(1978)}]{1978ApJ...221L..29B}
{Blandford}, R.~D. \& {Ostriker}, J.~P. 1978, \apjl, 221, L29

\bibitem[{{Cranmer} \& {van Ballegooijen}(2005)}]{2005ApJS..156..265C}
{Cranmer}, S.~R. \& {van Ballegooijen}, A.~A. 2005, \apjs, 156, 265

\bibitem[{{Krymskii}(1977)}]{1977DoSSR.234R1306K}
{Krymskii}, G.~F. 1977, Akademiia Nauk SSSR Doklady, 234, 1306

\bibitem[{Lax \& Wendroff(1960)}]{LW60}
Lax, P. \& Wendroff, B. 1960, Commun. Pure Appl. Math., 13, 217

\bibitem[{{Vainio}(2003)}]{2003A&A...406..735V}
{Vainio}, R. 2003, \aap, 406, 735

\bibitem[{{Vainio} \& {Laitinen}(2007)}]{2007ApJ...658..622V}
{Vainio}, R. \& {Laitinen}, T. 2007, \apj, 658, 622

\bibitem[{{Vainio} {et~al.}(2003){Vainio}, {Laitinen}, \&
  {Fichtner}}]{2003A&A...407..713V}
{Vainio}, R., {Laitinen}, T., \& {Fichtner}, H. 2003, \aap, 407, 713

\bibitem[{van Leer(1974)}]{vanLeer1974361}
van Leer, B. 1974, Journal of Computational Physics, 14, 361

\end{thebibliography}

\end{document}